\batchmode
\makeatletter
\def\input@path{{\string"E:/Trabajo Angel/Mis articulos/LOSPA fixed target number/\string"/}}
\makeatother
\documentclass[british,english]{IEEEtran}
\usepackage[T1]{fontenc}
\usepackage[latin9]{inputenc}
\usepackage{amsmath}
\usepackage{amssymb}

\makeatletter

\providecommand{\tabularnewline}{\\}

\usepackage{cite}
\allowdisplaybreaks 

\makeatother

\usepackage{babel}
\begin{document}

\title{Labelled OSPA metric for fixed and known number of targets}

\author{Ángel F. García-Fernández, Mark R. Morelande, Jesús Grajal\thanks{Ángel F. García-Fernández is with the Department of Electrical and Computer Engineering, Curtin University, Perth, WA 6102, Australia (e-mail: angel.garciafernandez@curtin.edu.au).

Mark R. Morelande is with the Department of Electrical and Electronic Engineering, The University of Melbourne, Parkville, Victoria 3010, Australia (email: mrmore@unimelb.edu.au)

Jesús Grajal is with Departamento de Se\~{n}ales, Sistemas y Radiocomunicaciones, Universidad Polit\'{e}cnica de Madrid, Ciudad Universitaria s/n, 28040 Madrid, Spain. (email: jesus@gmr.ssr.upm.es).
}}
\maketitle
\begin{abstract}
The evaluation of multiple target tracking algorithms with labelled
sets can be done using the labelled optimal subpattern assignment
(LOSPA) metric. In this paper, we provide the expression of the same
metric for fixed and known number of targets when vector notation
is used.\thispagestyle{empty}\end{abstract}


\section{Introduction}

Multitarget tracking systems should solve two basic problems. The
first one is to estimate the number of targets and their states at
the current time. The second one is to connect target state estimates
that belong to the same target along time to form tracks. The conventional
way of building tracks in the random finite set framework (RFS) \cite{Mahler_book07}
is to attach a label to the individual target states \foreignlanguage{british}{\cite{Ma06,Vo13}}.

Labels have two important properties: they are unique (no two targets
can have the same label) and they are fixed over time. Labels were
used for track formation in \cite{Angel13,Angel09} using a vector-based
formulation and in \cite{Ma06,Vo13} using the RFS framework. The
approaches of \cite{Angel13,Angel09} and \cite{Ma06,Vo13} are equivalent
due to the bijection between the labelled RFS state and the hybrid
labelled multitarget state vector \cite[Appendix B]{Angel_thesis11}.
For the same reason, for fixed and known number of targets, representing
the multitarget state as a vector is equivalent to a labelled set.
One way to evaluate performance of tracking algorithms based on labelled
set is using the labelled optimal subpattern assignment (LOSPA) metric
\cite{Ristic11}. 

In some cases, it is convenient to assume that the number of targets
is fixed and known \cite{Svensson11,Guerriero10}. This way we can
study some properties of tracking algorithms more easily. In these
cases, it is usually useful to use vector notation, in which labels
are implicit in the ordering of the components of a vector, to denote
a labelled set. The problem is that the LOSPA metric in \cite{Ristic11}
is defined with explicit labels. In this paper, we fill this gap and
provide an expression for this metric when the number of targets is
fixed and known and vector notation is used. 

This paper is organised as follows. In Section \ref{sec:Labelled-set-and},
we introduce the two equivalent representations of the multitarget
state based on a labelled set and a vector. We provide the expression
for the LOSPA metric using vector notation in Section \ref{sec:Labelled-OSPA-metric}.

\section{Labelled set and vector notation\label{sec:Labelled-set-and}}

In this paper, we make the following assumption
\begin{itemize}
\item A The number of targets is fixed and known
\end{itemize}
Under Assumption A, the labelled set that contains the targets and
the labels is represented as $\left\{ \left[\left(\mathbf{x}_{1}^{k}\right)^{T},l_{1}\right]^{T},\left[\left(\mathbf{x}_{2}^{k}\right)^{T},l_{2}\right]^{T},...,\left[\left(\mathbf{x}_{t}^{k}\right)^{T},l_{t}\right]^{T}\right\} $
where $l_{j}\in\mathbb{R}$ represents the $j$th label, $\mathbf{x}_{j}^{k}\in\mathbb{R}^{n_{x}}$
is the state vector at time $k$ for target with label $l_{j}$ and
$T$ denotes transpose. Labels are unique, assigned deterministically
and do not change with time. Therefore, the same information of the
labelled set $\left\{ \left[\left(\mathbf{x}_{1}^{k}\right)^{T},l_{1}\right]^{T},\left[\left(\mathbf{x}_{2}^{k}\right)^{T},l_{2}\right]^{T},...,\left[\left(\mathbf{x}_{t}^{k}\right)^{T},l_{t}\right]^{T}\right\} $
is contained in the multitarget state vector $\mathbf{X}^{k}=\left[\left(\mathbf{x}_{1}^{k}\right)^{T},\left(\mathbf{x}_{2}^{k}\right)^{T},...,\left(\mathbf{x}_{t}^{k}\right)^{T}\right]^{T}\in\mathbb{R}^{tn_{x}}$.
The labels of the labelled set are implicit in the ordering inherent
in the multitarget state vector components and we can establish a
bijection between the multitarget state vector and the labelled set.

Under Assumption A, it is usually more convenient to use the multitarget
state vector than the labelled set because we do not have to carry
along the explicit labels. This is for example useful when performing
Bayesian inference.

\section{Labelled OSPA metric with vector notation\label{sec:Labelled-OSPA-metric}}

Here, we provide the expression of the LOSPA metric for labelled sets
defined in \cite{Ristic11} when we use the vector notation under
Assumption A. We also prove it is a metric with this notation for
completeness. It should be noted that the triangle inequality for
LOSPA metric using labelled sets is proved in \cite{Ristic13}.

We represent the permutations of vector $\left[1,...,t\right]^{T}$
as vectors $\boldsymbol{\phi}_{i}=\left[\phi_{i,1},...,\phi_{i,t}\right]^{T}$
$i\in\left\{ 1,...,t!\right\} $. Then, the labelled OSPA (LOSPA)
distance between multitarget vectors $\mathbf{A}^{k}=\left[\left(\mathbf{a}_{1}^{k}\right)^{T},\left(\mathbf{a}_{2}^{k}\right)^{T},...,\left(\mathbf{a}_{t}^{k}\right)^{T}\right]^{T}\in\mathbb{R}^{tn_{x}}$
and $\mathbf{B}^{k}=\left[\left(\mathbf{b}_{1}^{k}\right)^{T},\left(\mathbf{b}_{2}^{k}\right)^{T},...,\left(\mathbf{b}_{t}^{k}\right)^{T}\right]^{T}\in\mathbb{R}^{tn_{x}}$
is
\begin{align}
 & d\left(\mathbf{A}^{k},\mathbf{B}^{k}\right)=\nonumber \\
 & \quad\left(\frac{1}{t}\min_{i\in\left\{ 1,...,t!\right\} }\left[\sum_{j=1}^{t}b^{p}\left(\mathbf{a}_{j}^{k},\mathbf{b}_{\phi_{i,j}}^{k}\right)+\alpha^{p}\overline{\delta}\left[j-\phi_{i,j}\right]\right]\right)^{1/p}\label{eq:LOSPA_vectors}
\end{align}
where $\overline{\delta}\left[\cdot\right]$ is the complement of
the Kronecker delta, i.e., $\overline{\delta}\left[j\right]=0$ if
$j=0$ and $\overline{\delta}\left[j\right]=1$ otherwise, $\alpha>0$,
$1\leq p<\infty$ and $b\left(\cdot,\cdot\right)$ is a metric on
the space $\mathbb{R}^{n_{x}}$. In \cite{Ristic11} the authors include
another parameter $p'$, we set $p'=p$ for simplicity. Function $d\left(\cdot,\cdot\right)$
is a metric as it satisfies the axioms of identity, symmetry and triangle
inequality. The identity and symmetry are straightforward to check.
The proof of the triangle inequality is given in Appendix \ref{sec:AppendA}.
 It should be noted that if $\alpha=0$, we get the optimal subpattern
assignment metric (OSPA) without cut-off distance \cite{Schuhmacher08}
and not the LOSPA. In Appendix \ref{sec:AppendC}, we prove that this
metric is equivalent in the labelled set domain.

\subsubsection*{Illustrative example\label{sub:Illustrative-example-LOSPA}}

We illustrate how the LOSPA metric works in a simple example. Let
us assume there are three unidimensional targets and the multitarget
state is $\mathbf{X}^{k}=[-10,0,10]^{T}$. That is, target 1 is at
-10, target 2 is at 0 and target 3 is at 10. We use the Euclidean
metric for $b\left(\cdot,\cdot\right)$ with $p=2$. The LOSPA between
$\mathbf{X}^{k}$ and several estimates $\hat{\mathbf{X}}^{k}$, which
only differ in their labelling, are given in Table \ref{tab:LOSPA_example}.
As all the estimates only differ in their labelling, they have the
same OSPA, which is 0.1. This implies that all the estimates have
the same accuracy as regards where the targets are. However, the first
estimate is closer in the LOSPA sense than the rest. The higher $\alpha$
is, the more the metric penalises wrong labelling/ordering. 

\begin{table}
\protect\caption{\selectlanguage{british}%
\label{tab:LOSPA_example}LOSPA between $\hat{\mathbf{X}}^{k}$ and
$\mathbf{X}^{k}=[-10,0,10]^{T}$\selectlanguage{english}%
}

\centering{}%
\begin{tabular}{ccc}
\hline 
Estimate $\hat{\mathbf{X}}^{k}$ &
LOSPA $\left(\alpha=0.1\right)$ &
LOSPA $\left(\alpha=1\right)$\tabularnewline
\hline 
$[-10.1,0.1,10.1]^{T}$ &
0.1 &
0.1\tabularnewline
$[0.1,-10.1,10.1]^{T}$ &
$\sqrt{0.1^{2}+0.02/3}$ &
$\sqrt{0.1^{2}+2/3}$\tabularnewline
$[10.1,-10.1,0.1]^{T}$ &
$\sqrt{0.1^{2}+0.03/3}$ &
$\sqrt{0.1^{2}+3/3}$\tabularnewline
\hline 
\end{tabular}
\end{table}

\appendices{}

\section{\label{sec:AppendA}}

In this appendix we prove the triangle inequality of the LOSPA metric,
which is given by (\ref{eq:LOSPA_vectors}). We want to show that
\begin{equation}
d\left(\mathbf{X}^{k},\mathbf{Y}^{k}\right)\leq d\left(\mathbf{X}^{k},\mathbf{Z}^{k}\right)+d\left(\mathbf{Z}^{k},\mathbf{Y}^{k}\right)
\end{equation}

\begin{flushleft}
As (\ref{eq:append1}) in Appendix \ref{sec:AppendB} is met for any
$m\in\left\{ 1,...,t!\right\} $ and $i\in\left\{ 1,...,t!\right\} $,
we can write
\par\end{flushleft}

\begin{align*}
 & \min_{i\in\left\{ 1,...,t!\right\} }\sqrt[p]{\sum_{j=1}^{t}b^{p}\left(\mathbf{x}_{j}^{k},\mathbf{y}_{\phi_{i,j}}^{k}\right)+\alpha^{p}\overline{\delta}\left[j-\phi_{i,j}\right]}\\
 & \qquad\leq\min_{m\in\left\{ 1,...,t!\right\} }\min_{i\in\left\{ 1,...,t!\right\} }\\
 & \qquad\quad\left[\sqrt[p]{\sum_{j=1}^{t}b^{p}\left(\mathbf{x}_{j}^{k},\mathbf{z}_{\phi_{m,j}}^{k}\right)+\alpha^{p}\overline{\delta}\left[j-\phi_{m,j}\right]}\right.\\
 & \qquad\quad\left.+\sqrt[p]{\sum_{j=1}^{t}b^{p}\left(\mathbf{z}_{\phi_{m,j}}^{k},\mathbf{y}_{\phi_{i,j}}^{k}\right)+\alpha^{p}\overline{\delta}\left[\phi_{m,j}-\phi_{i,j}\right]}\right]\\
 & \qquad=\min_{m\in\left\{ 1,...,t!\right\} }\left[\sqrt[p]{\sum_{j=1}^{t}b^{p}\left(\mathbf{x}_{j}^{k},\mathbf{z}_{\phi_{m,j}}^{k}\right)+\alpha^{p}\overline{\delta}\left[j-\phi_{m,j}\right]}\right.\\
 & \qquad\quad+\min_{i\in\left\{ 1,...,t!\right\} }\left.\sqrt[p]{\sum_{j=1}^{t}b^{p}\left(\mathbf{z}_{\phi_{m,j}}^{k},\mathbf{y}_{\phi_{i,j}}^{k}\right)+\alpha^{p}\overline{\delta}\left[\phi_{m,j}-\phi_{i,j}\right]}\right]
\end{align*}

By using a change of variables $\phi_{m,j}=j$ for $j=1...t$, we
get 
\begin{align*}
 & \min_{i\in\left\{ 1,...,t!\right\} }\left.\sqrt[p]{\sum_{j=1}^{t}b^{p}\left(\mathbf{z}_{\phi_{m,j}}^{k},\mathbf{y}_{\phi_{i,j}}^{k}\right)+\alpha^{p}\overline{\delta}\left[\phi_{m,j}-\phi_{i,j}\right]}\right]\\
 & =\min_{i\in\left\{ 1,...,t!\right\} }\left.\sqrt[p]{\sum_{j=1}^{t}b^{p}\left(\mathbf{z}_{j}^{k},\mathbf{y}_{\phi_{i,j}}^{k}\right)+\alpha^{p}\overline{\delta}\left[j-\phi_{i,j}\right]}\right]
\end{align*}
which is independent of $m$. Therefore, 
\begin{align*}
 & \min_{i\in\left\{ 1,...,t!\right\} }\sqrt[p]{\sum_{j=1}^{t}b^{p}\left(\mathbf{x}_{j}^{k},\mathbf{y}_{\phi_{i,j}}^{k}\right)+\alpha^{p}\overline{\delta}\left[j-\phi_{i,j}\right]}\\
 & \leq\min_{m\in\left\{ 1,...,t!\right\} }\left[\sqrt[p]{\sum_{j=1}^{t}b^{p}\left(\mathbf{x}_{j}^{k},\mathbf{z}_{\phi_{m,j}}^{k}\right)+\alpha^{p}\overline{\delta}\left[j-\phi_{m,j}\right]}\right]\\
 & \quad+\min_{i\in\left\{ 1,...,t!\right\} }\left.\sqrt[p]{\sum_{j=1}^{t}b^{p}\left(\mathbf{z}_{j}^{k},\mathbf{y}_{\phi_{i,j}}^{k}\right)+\alpha^{p}\overline{\delta}\left[j-\phi_{i,j}\right]}\right]
\end{align*}

Then, we can write 
\begin{align*}
 & \left(\frac{1}{t}\min_{i\in\left\{ 1,...,t!\right\} }\left[\sum_{j=1}^{t}b^{p}\left(\mathbf{x}_{j}^{k},\mathbf{y}_{\phi_{i,j}}^{k}\right)+\alpha^{p}\overline{\delta}\left[j-\phi_{i,j}\right]\right]\right)^{1/p}\\
 & \qquad\leq\left(\frac{1}{t}\min_{i\in\left\{ 1,...,t!\right\} }\left[\sum_{j=1}^{t}b^{p}\left(\mathbf{x}_{j}^{k},\mathbf{z}_{\phi_{i,j}}^{k}\right)+\alpha^{p}\overline{\delta}\left[j-\phi_{i,j}\right]\right]\right)^{1/p}\\
 & \qquad\quad+\left(\frac{1}{t}\min_{i\in\left\{ 1,...,t!\right\} }\left[\sum_{j=1}^{t}b^{p}\left(\mathbf{z}_{j}^{k},\mathbf{y}_{\phi_{i,j}}^{k}\right)+\alpha^{p}\overline{\delta}\left[j-\phi_{i,j}\right]\right]\right)^{1/p}
\end{align*}

Using (\ref{eq:LOSPA_vectors}), we complete the proof of the triangle
inequality.

\section{\label{sec:AppendB}}

This appendix provides a subsidiary result that is necessary for the
proof of the triangle inequality in Appendix \ref{sec:AppendA}. Using
the fact that $b\left(\cdot,\cdot\right)$ is a metric
\begin{align}
b\left(\mathbf{x}_{j}^{k},\mathbf{y}_{\phi_{i,j}}^{k}\right)\leq & b\left(\mathbf{x}_{j}^{k},\mathbf{z}_{\phi_{m,j}}^{k}\right)+b\left(\mathbf{z}_{\phi_{m,j}}^{k},\mathbf{y}_{\phi_{i,j}}^{k}\right)\nonumber \\
b^{p}\left(\mathbf{x}_{j}^{k},\mathbf{y}_{\phi_{i,j}}^{k}\right)\leq & \left(b\left(\mathbf{x}_{j}^{k},\mathbf{z}_{\phi_{m,j}}^{k}\right)+b\left(\mathbf{z}_{\phi_{m,j}}^{k},\mathbf{y}_{\phi_{i,j}}^{k}\right)\right)^{p}\label{eq:AppendixB_1}
\end{align}
In addition
\begin{align}
\alpha\overline{\delta}\left[j-\phi_{i,j}\right]\leq & \alpha\overline{\delta}\left[j-\phi_{m,j}\right]+\alpha\overline{\delta}\left[\phi_{m,j}-\phi_{i,j}\right]\nonumber \\
\left(\alpha\overline{\delta}\left[j-\phi_{i,j}\right]\right)^{p}\leq & \left(\alpha\overline{\delta}\left[j-\phi_{m,j}\right]+\alpha\overline{\delta}\left[\phi_{m,j}-\phi_{i,j}\right]\right)^{p}\label{eq:AppendixB_2}
\end{align}
Using (\ref{eq:AppendixB_1}) and (\ref{eq:AppendixB_2}), we get
\begin{align}
 & b^{p}\left(\mathbf{x}_{j}^{k},\mathbf{y}_{\phi_{i,j}}^{k}\right)+\left(\alpha\overline{\delta}\left[j-\phi_{i,j}\right]\right)^{p}\nonumber \\
 & \qquad\leq\left(b\left(\mathbf{x}_{j}^{k},\mathbf{z}_{\phi_{m,j}}^{k}\right)+b\left(\mathbf{z}_{\phi_{m,j}}^{k},\mathbf{y}_{\phi_{i,j}}^{k}\right)\right)^{p}\nonumber \\
 & \qquad\quad+\left(\alpha\overline{\delta}\left[j-\phi_{m,j}\right]+\alpha\overline{\delta}\left[\phi_{m,j}-\phi_{i,j}\right]\right)^{p}\\
 & \sum_{j=1}^{t}b^{p}\left(\mathbf{x}_{j}^{k},\mathbf{y}_{\phi_{i,j}}^{k}\right)+\left(\alpha\overline{\delta}\left[j-\phi_{i,j}\right]\right)^{p}\nonumber \\
 & \qquad\leq\sum_{j=1}^{t}\left(b\left(\mathbf{x}_{j}^{k},\mathbf{z}_{\phi_{m,j}}^{k}\right)+b\left(\mathbf{z}_{\phi_{m,j}}^{k},\mathbf{y}_{\phi_{i,j}}^{k}\right)\right)^{p}\nonumber \\
 & \qquad\quad+\left(\alpha\overline{\delta}\left[j-\phi_{m,j}\right]+\alpha\overline{\delta}\left[\phi_{m,j}-\phi_{i,j}\right]\right)^{p}\\
 & \left[\sum_{j=1}^{t}b^{p}\left(\mathbf{x}_{j}^{k},\mathbf{y}_{\phi_{i,j}}^{k}\right)+\left(\alpha\overline{\delta}\left[j-\phi_{i,j}\right]\right)^{p}\right]^{1/p}\nonumber \\
 & \qquad\leq\left[\sum_{j=1}^{t}\left(b\left(\mathbf{x}_{j}^{k},\mathbf{z}_{\phi_{m,j}}^{k}\right)+b\left(\mathbf{z}_{\phi_{m,j}}^{k},\mathbf{y}_{\phi_{i,j}}^{k}\right)\right)^{p}\right.\nonumber \\
 & \qquad\quad\left.+\left(\alpha\overline{\delta}\left[j-\phi_{m,j}\right]+\alpha\overline{\delta}\left[\phi_{m,j}-\phi_{i,j}\right]\right)^{p}\right]^{1/p}\label{eq:append4}
\end{align}

Using Minkowski inequality \cite{Hardy_book34}
\begin{align}
 & \left[\sum_{j=1}^{t}\left(b\left(\mathbf{x}_{j}^{k},\mathbf{z}_{\phi_{m,j}}^{k}\right)+b\left(\mathbf{z}_{\phi_{m,j}}^{k},\mathbf{y}_{\phi_{i,j}}^{k}\right)\right)^{p}\right.\nonumber \\
 & \left.+\left(\alpha\overline{\delta}\left[j-\phi_{m,j}\right]+\alpha\overline{\delta}\left[\phi_{m,j}-\phi_{i,j}\right]\right)^{p}\right]^{1/p}\nonumber \\
 & \qquad\leq\left[\sum_{j=1}^{t}b^{p}\left(\mathbf{x}_{j}^{k},\mathbf{z}_{\phi_{m,j}}^{k}\right)+\alpha^{p}\overline{\delta}\left[j-\phi_{m,j}\right]\right]^{1/p}\nonumber \\
 & \qquad\quad+\left[\sum_{j=1}^{t}b^{p}\left(\mathbf{z}_{\phi_{m,j}}^{k},\mathbf{y}_{\phi_{i,j}}^{k}\right)+\alpha^{p}\overline{\delta}\left[\phi_{m,j}-\phi_{i,j}\right]\right]^{1/p}\label{eq:minkowski}
\end{align}

Using (\ref{eq:minkowski}) into (\ref{eq:append4}), we get
\begin{align}
 & \left[\sum_{j=1}^{t}b^{p}\left(\mathbf{x}_{j}^{k},\mathbf{y}_{\phi_{i,j}}^{k}\right)+\alpha^{p}\overline{\delta}\left[j-\phi_{i,j}\right]\right]^{1/p}\nonumber \\
 & \qquad\leq\left[\sum_{j=1}^{t}b^{p}\left(\mathbf{x}_{j}^{k},\mathbf{z}_{\phi_{m,j}}^{k}\right)+\alpha^{p}\overline{\delta}\left[j-\phi_{m,j}\right]\right]^{1/p}\nonumber \\
 & \qquad\quad\left[\sum_{j=1}^{t}b^{p}\left(\mathbf{z}_{\phi_{m,j}}^{k},\mathbf{y}_{\phi_{i,j}}^{k}\right)+\alpha^{p}\overline{\delta}\left[\phi_{m,j}-\phi_{i,j}\right]\right]^{1/p}\label{eq:append1}
\end{align}
for any $m\in\left\{ 1,...,t!\right\} $.

\section{\label{sec:AppendC}}

In this appendix we show that the metric used in this paper is equivalent
in the set domain under Assumption A. The labelled OSPA metric $d_{s}\left(\cdot,\cdot\right)$
in the set domain \cite{Ristic11} requires the definition of the
labelled sets
\begin{align*}
A^{k}= & \left\{ \left[\left(\mathbf{a}_{1}^{k}\right)^{T},l_{1}\right]^{T},\left[\left(\mathbf{a}_{2}^{k}\right)^{T},l_{2}\right]^{T},...,\left[\left(\mathbf{a}_{t}^{k}\right)^{T},l_{t}\right]^{T}\right\} \\
B^{k}= & \left\{ \left[\left(\mathbf{b}_{1}^{k}\right)^{T},l_{1}\right]^{T},\left[\left(\mathbf{b}_{2}^{k}\right)^{T},l_{2}\right]^{T},...,\left[\left(\mathbf{b}_{t}^{k}\right)^{T},l_{t}\right]^{T}\right\} 
\end{align*}
 where $l_{1},...,l_{t}$ are the explicit labels of the targets that
must be used in the set approach. Then,
\begin{align*}
 & d_{s}\left(A^{k},B^{k}\right)\\
 & \qquad=\left(\frac{1}{t}\min_{i\in\left\{ 1,...,t!\right\} }\left[\sum_{j=1}^{t}b^{p}\left(\mathbf{a}_{j}^{k},\mathbf{b}_{\phi_{i,j}}^{k}\right)+\alpha^{p}\overline{\delta}\left[l_{j}-l_{\phi_{i,j}}\right]\right]\right)^{1/p}\\
 & \qquad=\left(\frac{1}{t}\min_{i\in\left\{ 1,...,t!\right\} }\left[\sum_{j=1}^{t}b^{p}\left(\mathbf{a}_{j}^{k},\mathbf{b}_{\phi_{i,j}}^{k}\right)+\alpha^{p}\overline{\delta}\left[j-\phi_{i,j}\right]\right]\right)^{1/p}\\
 & \qquad=d\left(\mathbf{A}^{k},\mathbf{B}^{k}\right)
\end{align*}

\bibliographystyle{IEEEtran}
\bibliography{0E__Trabajo_Angel_Mis_articulos_LOSPA_fixed_target_number_Referencias}

\end{document}